\documentstyle[aps,epsfig,floats]{revtex}

\tighten
\newcommand{\beq}{\begin{equation}}
\newcommand{\eeq}{\end{equation}}
\newcommand{\ba}{\begin{eqnarray}}
\newcommand{\ea}{\end{eqnarray}}

\newcommand{\xbj}{x_{\scriptscriptstyle B}}
\newcommand{\bpt}{\bm p_\st}
\newcommand{\bkt}{\bm k_\st}
\newcommand{\psibar}{\overline{\psi}}
\newcommand{\la}{\langle}
\newcommand{\ra}{\rangle}
\newcommand{\amp}[1]{\la #1 \ra}
\newcommand{\Tr}{{\text{Tr}}}

\newcommand{\slsh}[1]{\mbox{$\not\! #1$}}
\newcommand{\bm}[1]{\bbox{#1}}

\newcommand{\st}{{\scriptscriptstyle T}}
\newcommand{\sL}{{\scriptscriptstyle L}}
\newcommand{\be}{\begin{equation}}
\newcommand{\ee}{\end{equation}}
\newcommand{\bea}{\begin{eqnarray}}
\newcommand{\eea}{\end{eqnarray}}
\newcommand{\zh}{z_h}

\begin{document}
 
\draft
\title{
\begin{flushright}
\begin{minipage}{4 cm}
\small
hep-ph/0104271
\end{minipage}
\end{flushright}
Evolution of transverse momentum dependent distribution \\
and fragmentation functions}

\author{
A.A. Henneman$^1$, 
Dani\"el Boer$^2$, 
and P.J.\ Mulders$^1$}
\address{\mbox{}\\
$^1$Division of Physics and Astronomy, Vrije Universiteit \\
De Boelelaan 1081, NL-1081 HV Amsterdam, the Netherlands
\mbox{}\\
$^2$RIKEN-BNL Research Center\\
Brookhaven National Laboratory, Upton, NY 11973, U.S.A.\\
}

\maketitle

\begin{abstract}
We use Lorentz invariance and the QCD equations of motion to study
the evolution of functions that appear at leading order in a $1/Q$
expansion in azimuthal asymmetries. 
This includes the evolution equation of the Collins fragmentation function. 
The moments of these functions are matrix elements of known twist two 
and twist three operators.
We present the evolution in the large $N_c$ limit, restricting to non-singlet 
for the chiral-even functions.
\end{abstract}

\pacs{}  

\section{Introduction}

Azimuthal asymmetries in hard scattering processes with at least two relevant
hadrons constitute a rich phenomenology, suitable for studying quark and gluon
correlations in hadrons. By relevant hadrons we mean hadrons used as target
or detected in the final state. A well-known azimuthal asymmetry appears in 
the semi-inclusive deep inelastic polarized leptoproduction of pions
($e p^\uparrow \to e' \pi X$) generated by the so-called Collins 
effect~\cite{Collins-93b}. This asymmetry is one of
the few possibilities to gain access to the so-called transversity or
transverse spin distribution function \cite{RS-79,Jaffe-Ji-92}, which is
the third distribution function needed for the complete characterization of the
(collinear) spin state of a proton as probed in hard scattering processes. 
In contrast to the transversity function, the evolution of the Collins 
fragmentation function has not been presented yet. Knowledge of this evolution
is indispensable for relating measurements at different energies. 
              
In processes like the semi-inclusive leptoproduction mentioned above, it is 
important to take transverse momentum of partons into account. The parton 
distribution functions as 
a function of a light-cone momentum fraction and transverse momentum have 
first been studied by Ralston and Soper~\cite{RS-79} for the Drell-Yan
process at tree level. The precise operator definition of such transverse 
momentum ($p_\st$) dependent 
distribution (and also fragmentation) functions is a non-trivial
issue (mainly because of gauge invariance) and in several studies this 
matter has been addressed \cite{CS81,CS82,CSS83,Boer-Mulders-00}. Besides 
their definition and appearance in cross sections, they have not yet 
been considered beyond tree level. One thing that one wants to know is how 
these $p_\st$ dependent distribution and fragmentation functions evolve, for
instance the fragmentation function pointed out by Collins, which is one of
the new functions entering in the description of hadrons when transverse
momentum is considered. Its evolution will be one of the new
results presented in this paper, although we limit ourselves to the large
$N_c$ limit, in which case the evolution is an autonomous one. 
 
In general, factorization of hard scattering processes means that it is
possible to separate parts containing only soft or hard physics. In the
perturbative calculation of the hard subprocess (the partonic cross section)
one encounters collinear divergences, which one can absorb into the matrix
elements (or equivalently the distribution or fragmentation functions) 
describing the soft part of the process. This redefinition of the soft parts 
should be possible to all orders in the coupling constant. This procedure 
introduces a factorization scale and the goal is to calculate the 
dependence on this scale, which determines the high-energy behavior of the 
cross section. 

Factorization crucially depends on the presence of a large energy scale in 
the process, such as the space-like momentum transfer squared in 
leptoproduction or the time-like momentum squared of a lepton pair in
Drell-Yan scattering. In this paper we will be concerned with functions that
appear in processes which have, apart from such a hard scale, an
additional soft momentum scale, related to the transverse momentum 
of the partons. The first factorization theorem for such a situation was
obtained in Ref.\ \cite{CS81} for the process $e^+ e^- \to h_1 h_2 X$, where
the vector boson has a large invariant mass $Q$, but a small transverse
momentum $Q_\st$ 
with respect to the two almost back-to-back hadrons $h_1$ and $h_2$, i.e.\
$Q_\st^2 \ll Q^2$. 
Similar situations occur in the Drell-Yan process, where one 
has besides the momentum of the lepton pair, two hadron momenta and 
in one-hadron inclusive leptoproduction where one also deals
with three momenta: the large momentum transfer, the target momentum and the
momentum of the produced hadron. 

The effects of parton transverse momenta lead to azimuthal asymmetries
in such processes, often coupled to the spin of the partons and/or hadrons. 
Just as for spin asymmetries, the azimuthal asymmetries provide a rich new 
phenomenology in Drell-Yan scattering, semi-inclusive deep inelastic
scattering and electron-positron 
annihilation~\cite{RS-79,LM94,Kotzinian95,MT96,Boer-Mulders-98,Boer-J-M-99,BJM-97}.
In this paper we will study the scale dependence of the various distribution 
and fragmentation functions appearing in these (polarized) processes.  
We do
this for specific moments in both $p_\st$ and $x$, employing Lorentz 
invariance and the
QCD equations of motion. The moments in $x$ for leading (collinear) 
distribution functions
(appearing for instance in inclusive leptoproduction) are related to matrix 
elements of twist two operators. On the other hand, for the transverse 
moments entering the azimuthal asymmetry expressions of interest, one finds 
relations to matrix elements of twist two {\em and\/} twist three operators, 
for which the evolution, however, is known. In the large $N_c$ limit this 
evolution becomes particularly simple and is known from studies of inclusive 
processes where the transverse moments can be eliminated from the expressions. 
This knowledge allows us to obtain the evolution
equations for the desired $p_\st$ and $x$ moments of the transverse momentum
dependent distribution and fragmentation functions, that enter in azimuthal
(spin) asymmetries. Such asymmetries recently have gained in interest, as can
be seen from the experimental studies in 
Refs.\ \cite{Efremov,Bravar-99,HERMES}.  

\section{Formalism}

In this paper we will study the scale dependence of the distribution
and fragmentation functions in (for example) 
one-hadron inclusive leptoproduction ($eH\rightarrow ehX$)
at leading order in an expansion in $1/Q$, where $q^2 = -Q^2$ is the
space-like momentum transfer squared. Experimentally, we are interested
in azimuthal asymmetries in the current fragmentation region, in which
case the target hadron momentum $P$ and the produced hadron momentum
$P_h$ satisfy $P\cdot P_h \sim Q^2$. We introduce two light-like vectors
via the hadron momenta and parameterize 
\bea
&&
P = \frac{\xi\,M^2}{\tilde Q\sqrt{2}}\,n_- +
\frac{\tilde Q}{\xi \sqrt{2}}\,n_+ ,
\\ &&
P_h = \frac{\zeta\,\tilde Q}{\sqrt{2}}\,n_- +
\frac{M_h^2}{\zeta \tilde Q\sqrt{2}}\,n_+,
\\ &&
q = \frac{\tilde Q}{\sqrt{2}}\,n_-
- \frac{\tilde Q}{\sqrt{2}}\,n_+ + q_\st ,
\eea
where $n_+$ and $n_-$ are two light-like vectors, chosen such that 
$n_+\cdot n_-=1$. We will often refer to the $\pm$ components of a vector $p$,
which are defined as $p^\pm=p\cdot n_\mp$.  We define the transverse
momentum scale $Q_\st^2 = - q_\st^2$. We are interested in the region where
$Q_\st^2 \ll Q^2$. Up to mass corrections of order $1/Q^2$ one then has
$\tilde Q^2  = Q^2 - Q_\st^2 \approx Q^2$. The ratio $\xi = -q^+/P^+ \approx
Q^2/2P\cdot q = \xbj$ is the Bjorken scaling variable and the ratio $\zeta =
P_h^-/q^-  \approx P\cdot P_h/P\cdot q = \zh$ is the usual fragmentation
variable.

In the case of inclusive deep inelastic scattering the soft part of the 
process is described by a correlation function. To be more specific, in 
leading order in powers of $1/Q$, the 
forward scattering Compton amplitude $T^{\mu\nu}$ can be written as 
\beq
T^{\mu\nu} = \int dx \, \Tr \left[ S^{\mu\nu} (x) \Phi (x) \right] + 
\cdots
\eeq
Here, $S^{\mu\nu}$ and $\Phi$ are the hard and 
soft scattering parts, respectively. The hard part $S^{\mu\nu}$ is the
$\gamma^*$--quark forward scattering amplitude. 
The color gauge invariant soft part is defined as
\ba
\Phi_{ij} (x) &\equiv& \left. \int \frac{d \xi^-}{2\pi}\ e^{i\,p\cdot
\xi}\amp{P,S|\, \psibar_j (0) \,{\cal U}(0,\xi)
\,\psi_i(\xi) | P,S}\right|_{LC} .
\label{PhiDIS}
\ea
where the subscript `LC' indicates $\xi^+ = \xi_\st = 0$ and
\be
{\cal U}(0,\xi) =
{\cal P} \exp \left(-ig\int_{0}^{\xi^-} d \eta^- \, A^+(\eta) \right) ,
\ee
is a gauge link with the path running along $n_-$.
The correlation function $\Phi$ is a function of the light-cone momentum 
fraction $x = p^+/P^+$ only. The most general 
parameterization of $\Phi$ which is in accordance with 
the required symmetries (hermiticity, parity, time reversal) and
which is relevant for calculations at leading order in $1/Q$ is
given by
\be
\Phi^{{\rm twist}-2}(x)=\frac{1}{2} \left\{{f_1(x)} \slsh n_+ 
+ S_\sL\,{g_1(x)}\,\gamma_{5}\slsh n_+
+ {h_1(x)}\,\gamma_{5}\slsh S_\st\slsh n_+ \right\} ,
\ee
where also for the spin vector a decomposition in $n_\pm$ is adopted,
$S = S_\sL\,(P^+/M)\,n_+ - S_\sL\,(P^-/M)\,n_- + S_\st$.
Specifying also the flavor one also encounters the notations $q(x) =
f_1^q(x)$,  $\Delta q(x) = g_1^q(x)$ and $\delta q(x) = \Delta_T q(x) =
h_1^q(x)$. The evolution equations for these functions are
known to next-to-leading order \cite{Mertig96,Haya97}
and for the singlet $f_1$ and $g_1$ there is mixing with the unpolarized and
polarized gluon distribution functions $g(x)$ and $\Delta g(x)$, respectively. 

Denoting these functions as twist-2 makes sense
because the local operators connected to the Mellin moments of these functions
are related to the matrix elements of local twist-2 operators, like 
$\overline \psi \,\gamma^+(D^+)^n\,\psi$.

If one calculates $T_{\mu\nu}$ up to order $1/Q$ one needs also the 
$M/P^+$ parts in the parameterization of $\Phi(x)$
\cite{Jaffe-Ji-92,Boer-Mulders-98}
\bea
\Phi^{{\rm twist}-3}(x) &=& \frac{M}{2P^+} \left\{{e(x)} 
+ {g_T(x)}\,\gamma_{5}\slsh S_\st
+ S_\sL\,{h_L(x)}\,\gamma_{5}\frac{[\slsh n_+,\slsh n_-]}{2} \right\} 
\nonumber \\ 
& + & \frac{M}{2P^+} \left\{-i\,S_\sL\,{e_L(x)} \gamma_5
- {f_T(x)}\,\epsilon_\st^{\rho\sigma}\gamma_\rho S_{\st\sigma}
+ i\,{h(x)}\frac{[\slsh n_+,\slsh n_-]}{2} \right\}. 
\eea
We have not imposed time-reversal invariance 
in order to study also the T-odd functions,
which are important e.g.\ in the study of fragmentation functions. 
The functions $e$, $g_T$ and $h_L$ are T-even, the functions $e_L$,
$f_T$ and $h$ are T-odd (we will not concern ourselves with the formal 
problems related to T-odd distribution functions 
\cite{Collins-93b,Boer-Mulders-98}). 
The leading order evolution of $e$, $g_T$ and $h_L$
is known \cite{Bukh,Evol,Koike-95,Belitsky-Muller-97} and for the non-singlet
case this also provides the evolution of the T-odd functions $e_L$,
$f_T$ and $h$ respectively, for which the operators involved differ only from
those of the T-even functions by a $\gamma_5$ matrix. 

The twist assignments of these functions is better seen by considering the
light-cone correlators
\bea
\Phi^\alpha_{D\,ij}(x,y) & \equiv &
\left. \frac{P^+}{M}\int \frac{d\xi^-}{2\pi}\,\frac{d\eta^-}{2\pi}
\ e^{i\,p_1\cdot (\xi-\eta) + ip\cdot \eta} \langle P,S \vert 
\overline \psi_j(0)\,{\cal U}(0,\eta)\,iD_\st^\alpha (\eta)\,{\cal
U}(\eta,\xi)\,\psi_i(\xi) \vert P,S \rangle \right|_{LC},
\eea
depending on $x = p^+/P^+$ and $y = p_1^+/P^+$. It is parameterized in terms of
two-argument functions \cite{Jaffe-Ji-92}
\bea
\Phi_D^\alpha(x,y) &=& \frac{1}{2}\Biggl\{
G_D(x,y)\,i\epsilon_\st^{\alpha\beta}S_{\st\beta}\slsh n_+
+ \tilde G_D(x,y)\,S_\st^\alpha\,\gamma_5\slsh n_+
\nonumber \\ && \qquad \quad \mbox{}
+ H_D(x,y)\,S_\sL\,\gamma_5\gamma_\st^\alpha\slsh n_+
+ E_D(x,y)\,\gamma_\st^\alpha\slsh n_+\Biggr\},
\label{twoarg}
\eea
where parity invariance has been imposed.
Hermiticity leads to
$G_D^\ast(x,y) = - G_D(y,x)$,
$\tilde G_D^\ast(x,y) = \tilde G_D(y,x)$,
$H_D^\ast(x,y) = H_D(y,x)$,
and $E_D^\ast(x,y) = - E_D(y,x)$.
Time reversal invariance would require these functions to be real. 
The QCD equations of motion can be used to relate the twist-3 functions
appearing in the parameterization of $\Phi$ to the correlators
$\Phi_D^\alpha$. This gives the relations
\bea
&&
\int dy\,\left[G_D(x,y) + \tilde G_D(x,y)\right]= x\,g_T(x)
-\frac{m}{M}\,h_1(x) + ix\,f_T(x),
\\ &&
2\int dy\ H_D(x,y)=x\,h_L(x) - \frac{m}{M}\,g_1(x) - ix\,e_L(x),
\\ &&
2\int dy\ E_D(x,y)=x\,e(x) - \frac{m}{M}\,f_1(x) + ix\,h(x) .
\eea
The local operator matrix elements corresponding with the moments of
the functions in $\Phi_D^\alpha(x)$ are (note that $\alpha$ is transverse)
twist-3 operators, up to quark mass contributions multiplying twist-2
operators. 

\section{Transverse momentum dependent distribution functions}

If one considers a semi-inclusive hard scattering process in which two 
hadrons are identified (in either initial and/or final state), then
the treatment of transverse momentum is important. One needs to study 
correlation functions that also depend on the transverse momentum, 
$\Phi(x,\bm p_\st)$, for which the
most general parameterization involves more
functions. To be more specific, one needs the lightfront correlation function
\be
\Phi_{ij}(x,\bm{p}_\st) \equiv 
\left. \int \frac{d \xi^-\,d^2\xi_\st}{(2\pi)^3}\ e^{i\,p\cdot
\xi}\amp{P,S|\,\psibar_j (0)\,{\cal U}(0,-\infty)
\,{\cal U}(-\infty,\xi)\,\psi_i(\xi) | P,S}\right|_{\xi^+ = 0} .
\ee
At leading orders in powers of $1/Q$ the following transverse 
momentum dependent
distribution functions are needed to parameterize this correlation function
\cite{RS-79,Boer-Mulders-98} 
\begin{eqnarray}
\Phi(x,\bm{p}_\st) & = & 
\frac{1}{2}\,\Biggl\{
f_1(x ,\bm p^2_\st)\,\slsh n_+ + 
f_{1T}^\perp(x ,\bm p^2_\st)\, \frac{\epsilon_{\mu \nu \rho \sigma}\gamma^\mu 
n_+^\nu p_\st^\rho S_{\st}^\sigma}{M}
\nonumber \\ && \mbox{}  
- g_{1s}(x ,\bm p_\st)\, \slsh n_+ \gamma_5
- h_{1T}(x ,\bm p^2_\st)\,i\sigma_{\mu\nu}\gamma_5 S_{\st}^\mu n_+^\nu
\nonumber \\ && \mbox{}
- h_{1s}^\perp(x ,\bm p_\st)\,\frac{i\sigma_{\mu\nu}\gamma_5 p_\st^\mu
n_+^\nu}{M} + h_1^\perp (x,\bm p^2_\st) \, \frac{\sigma_{\mu\nu} p_\st^\mu
n_+^\nu}{M}\Biggl\}.
\label{paramPhixkt}
\end{eqnarray}
We used the shorthand notation
\beq
g_{1s}(x, \bm p_\st) \equiv
S_\sL\,g_{1L}(x ,\bm p_\st^2)
+ \frac{(\bpt\cdot\bm{S}_{\st})}{M}\,g_{1T}(x ,\bm p_\st^2), 
\eeq
and similarly for $h_{1s}^\perp$.
The parameterization contains two T-odd functions, the
Sivers function $f_{1T}^\perp$ \cite{s90,Anselmino} and the 
function $h_1^\perp$, the distribution function analogue of the 
Collins fragmentation function~\cite{Collins-93b}. 
Upon integration over $\bpt$, Eq.~(\ref{paramPhixkt})  
reduces to $\Phi(x)$ with
\ba
f_1(x) & \equiv & \int d^2 p_\st \, f_{1}(x,\bpt^2),\label{f1x}\\
g_1(x) & \equiv & \int d^2 p_\st \, g_{1L}(x,\bpt^2),\label{g1x}\\
h_1(x) & \equiv & \int d^2 p_\st \, \left[ h_{1T}(x,\bpt^2)
+ \frac{\bpt^2}{2M^2}\, h_{1T}^\perp (x,\bpt^2) \right].
\ea
The deep inelastic scattering process is only sensitive to the latter three
functions, but in semi-inclusive deep inelastic scattering or in 
the Drell-Yan process (at small $q_\st$), one is sensitive to
the $p_\st$-dependent functions.
At measured $q_\st$ one deals with a convolution of two $p_\st$-dependent
functions, where the transverse momenta of the partons from different
hadrons combine to $q_\st$~\cite{RS-79,MT96,Boer-00}. 
A decoupling is achieved by studying cross sections weighted with the
momentum $q_\st^\alpha$, leaving only the directional (azimuthal) dependence.
The functions that appear in that case are contained in
\be
\Phi_\partial^\alpha (x) \equiv
\int d^2 p_\st\,\frac{p_\st^\alpha}{M} \,\Phi(x,\bm p_\st)
\ee
which projects out the functions in $\Phi(x,\bm p_\st)$ where $p_\st$
appears linearly,
\bea
\Phi_\partial^\alpha (x) & = &
\frac{1}{2}\,\Biggl\{
-g_{1T}^{(1)}(x)\,S_\st^\alpha\,\slsh n_+\gamma_5
-S_\sL\,h_{1L}^{\perp (1)}(x)
\,\frac{[\gamma^\alpha,\slsh n_+]\gamma_5}{2}
\nonumber \\
&&\quad \mbox{}
-{f_{1T}^{\perp (1)}}(x)
\,\epsilon^{\alpha}_{\ \ \mu\nu\rho}\gamma^\mu n_-^\nu {S_\st^\rho}
- i\,{h_1^{\perp (1)}}(x)
\,\frac{[\gamma^\alpha, \slsh n_+]}{2}\Biggr\},
\label{Phid}
\eea
where we define $\bm p_\st^2/2M^2$-moments (transverse moments) as
\be
f^{(n)}(x) = \int d^2p_\st\,\left(\frac{\bm p_\st^2}{2M^2}\right)^n
\,f(x,\bm p_\st).
\label{pt2mom}
\ee
The functions
$h_1^\perp$ and $f_{1T}^\perp$ are T-odd, vanishing if T-reversal
invariance can be applied to the matrix element.

At this point one can invoke Lorentz invariance as a possibility to
rewrite some functions. All functions in $\Phi(x)$ and 
$\Phi_\partial^\alpha(x)$ involve nonlocal matrix elements of two quark 
fields. Before constraining
the matrix elements to the light-cone or lightfront only a limited number
of amplitudes can be written down~\cite{MT96}. This leads to the following
Lorentz-invariance relations~\cite{Bukh,MT96,Boer-Mulders-98}
\bea
&&g_T  = g_1 + \frac{d}{dx}\,g_{1T}^{(1)},
\label{gTrel}
\\
&&h_L = h_1 - \frac{d}{dx}\,h_{1L}^{\perp (1)},
\label{hLrel}
\\
&&f_T =  - \frac{d}{dx}\,f_{1T}^{\perp (1)},
\\
&&h =  - \frac{d}{dx}\,h_{1}^{\perp (1)}.
\label{rel4}
\eea
From these relations, it is clear that  
the $\bm p_\st^2/2M^2$ moments of the $p_\st$-dependent 
functions, appearing in $\Phi_\partial^\alpha(x)$, involve both twist-2
and twist-3 operators. 

Starting from the defining expression of $\Phi$, one obtains, after weighting 
with $p_\st$, the gauge 
invariant operator structure,
\bea
M(\Phi_\partial^{\alpha})_{ij}(x) & = & 
\int \frac{d\xi^-}{2\pi}\ e^{ip\cdot \xi}
\langle P,S\vert i\partial_\st^\alpha\left[\,\overline \psi_j(0)
\,{\cal U}(0,-\infty){\cal U}(-\infty,\xi)
\psi_i(\xi)\right] \vert P,S\rangle \biggr|_{\xi^+=\xi_\st=0}
\nonumber \\ & = &
\int \frac{d\xi^-}{2\pi}\ e^{ip\cdot \xi}
\biggl\{ \langle P,S\vert \overline \psi_j(0)
\,{\cal U}(0,\xi)\, iD_\st^\alpha\psi_i(\xi) \vert P,S\rangle
\biggr|_{\xi^+=\xi_\st=0}
\nonumber \\ && \mbox{}
- \langle P,S\vert \overline \psi_j(0)\,{\cal U}(0,-\infty)
\int_{-\infty}^{\xi^-}d\eta^- \,{\cal U}(-\infty,\eta)
\,gF^{+\alpha}(\eta)
\,{\cal U}(\eta,\xi)\,\psi_i(\xi) \vert P,S\rangle \biggr|_{\xi^+=\xi_\st=0}
\biggr\}.
\eea
To see the partonic interpretation of the second term, consider the light-cone
gauge ($A^+ = 0$), in which case $F^{+\alpha} = \partial_-A_\st^\alpha$ and
the gauge links become unity. Up to some for our purposes
not relevant complications with boundary terms, the expression thus reduces to
\be
\Phi_\partial^\alpha (x) = \int dy\ \left[\Phi_D^\alpha(x,y)
- \Phi_A^\alpha(x,y)\right],
\label{defpartial}
\ee
where the leading part of the $\Phi_A^\alpha$ matrix element (in gauge
$A^+$ = 0) is built from $\psi_+$ and $A_\st^\alpha$ fields. The correlator 
$\Phi_A^\alpha$
can be parameterized analogous to $\Phi_D^\alpha$ with
(interaction-dependent) functions $G_A$, $\tilde G_A$, $H_A$ and $E_A$ with
similar hermiticity properties as the functions in $\Phi_D^\alpha$. Using
Eq.~(\ref{defpartial}) we define the following combinations,
\bea
&&
\int dy\,\left[G_A(x,y) \!+\! \tilde G_A(x,y)\right]=
x\,g_T(x) - \frac{m}{M}\,h_1(x)-g_{1T}^{(1)}(x) 
+ i\left[ x\,f_T(x) \!+\! f_{1T}^{\perp(1)}(x)\right]
\equiv x\,\tilde g_T(x)
+ ix\,\tilde f_T(x),
\label{collinear1}
\\ &&
2\int dy\ H_A(x,y)=x\,h_L(x) - \frac{m}{M}\,g_1(x) 
+2\,h_{1L}^{\perp (1)}(x) - ix\,e_L(x)\equiv
x\,\tilde h_L(x) - ix\,\tilde e_L(x),
\\ &&
2\int dy\ E_A(x,y)=x\,e(x) - \frac{m}{M}\,f_1(x) + i\left[
x\,h(x) +2\,h_1^{\perp (1)}(x)\right]\equiv
x\,\tilde e(x) + ix\,\tilde h(x).
\label{collinear4}
\eea
In principle, one can connect the functions defined here to those appearing in
the treatments of Ellis, Furmanski and Petronzio~\cite{EFP-83} or to those in
the treatment of Jaffe and Soldate~\cite{Soldate}. We end this section by 
giving the
relation to the functions used in a more recent treatment by
Belitsky~\cite{Belitsky-97a,Belitsky-97b}; comparison of the equations of
motion Eqs.~(\ref{collinear1}) - (\ref{collinear4}) and Lorentz invariance 
relations Eqs.~(\ref{gTrel}) - (\ref{rel4}) with those given in
Ref.~\cite{Belitsky-97b} leads us to identify
\ba
g_{1T}^{(1)}(x) & = & \bar{K}(x) - \int_x^1 dy f(y),\\
x \tilde{g}_T (x) & = & \int dx' \bar{D}(x,x') + \int_x^1 dy f(y),
\ea
with $\bar{K}, \bar{D}$ as defined in Ref.~\cite{Belitsky-97b} and 
\beq
f(y) = \int dx' \frac{\bar{D}(y,x')+\bar{D}(x',y)}{x'-y}.
\eeq

\section{Relations between twist-3 functions and transverse moments}

Using the equations of
motion relations in Eqs.~(\ref{collinear1}) - (\ref{collinear4}) and the
relations based on Lorentz invariance in Eqs.~(\ref{gTrel}) - (\ref{rel4}),
it is straightforward to relate the various twist-3 functions and the
$\bm p_\st^2/2M^2$ (transverse) moments of $p_\st$-dependent
functions. The results, grouping relevant combinations, are 
\bea
&&
g_T(x) = 
\int_x^1 dy\ \frac{g_1(y)}{y}
+ \frac{m}{M}\left[ \frac{h_1(x)}{x} - \int_x^1 dy\ \frac{h_1(y)}{y^2}\right]
+ \left[ \tilde g_T(x) - \int_x^1 dy\ \frac{\tilde g_T(y)}{y}\right],
\\ &&
\frac{g_{1T}^{(1)}(x)}{x} =
\int_x^1 dy\ \frac{g_1(y)}{y}
- \frac{m}{M}\int_x^1 dy\ \frac{h_1(y)}{y^2}
- \int_x^1 dy\ \frac{\tilde g_T(y)}{y},
\\[0.5cm] &&
h_L(x) = 
2x\int_x^1 dy\ \frac{h_1(y)}{y^2}
+ \frac{m}{M}\left[ \frac{g_1(x)}{x} - 2x\int_x^1 dy\ \frac{g_1(y)}{y^3}\right]
+ \left[ \tilde h_L(x) - 2x\int_x^1 dy\ \frac{\tilde h_L(y)}{y^2}\right],
\\ &&
\frac{h_{1L}^{\perp(1)}(x)}{x^2} =
-\int_x^1 dy\ \frac{h_1(y)}{y^2}
+ \frac{m}{M}\int_x^1 dy\ \frac{g_1(y)}{y^3}
+ \int_x^1 dy\ \frac{\tilde h_L(y)}{y^2},
\\[0.5cm] &&
e(x) =
\tilde e(x) + \frac{m}{M}\,\frac{f_1(x)}{x},
\\[0.5cm] &&
f_T(x) = 
\left[ \tilde f_T(x) - \int_x^1 dy\ \frac{\tilde f_T(y)}{y}\right] ,
\\ &&
\frac{f_{1T}^{\perp (1)}(x)}{x} =
\int_x^1 dy\ \frac{\tilde f_T(y)}{y},
\\[0.5cm] &&
h(x) = 
\left[ \tilde h(x) - 2x\int_x^1 dy\ \frac{\tilde h(y)}{y^2}\right],
\\ &&
\frac{h_{1}^{\perp(1)}(x)}{x^2} =
\int_x^1 dy\ \frac{\tilde h(y)}{y^2} ,
\\[0.5cm] &&
e_L(x) =
\tilde e_L(x).
\eea
Note that often the combinations of tilde
functions between brackets are denoted by a single `interaction-dependent' 
function.

In order to study the evolution of these functions, we consider the moments
$[f]_n \equiv \int dx\ x^{n-1}\,f(x)$, giving
\bea
&&
[g_T]_n = \frac{1}{n}\,[g_1]_n 
+ \frac{n-1}{n}\,[\tilde g_T]_n
+ \frac{m}{M}\,\frac{n-1}{n}\,[h_1]_{n-1} \ ,
\\ &&
[g_{1T}^{(1)}]_n = \frac{1}{n+1}\left(
[g_1]_{n+1} - [\tilde g_T]_{n+1} - \frac{m}{M}\,[h_1]_n\right) \ ,
\\[0.5cm] &&
[h_L]_n = \frac{2}{n+1}\,[h_1]_n 
+ \frac{n-1}{n+1}\,[\tilde h_L]_n
+ \frac{m}{M}\,\frac{n-1}{n+1}\,[g_1]_{n-1} \ ,
\\ &&
[h_{1L}^{\perp (1)}]_n = - \frac{1}{n+2}\left(
[h_1]_{n+1} - [\tilde h_L]_{n+1} - \frac{m}{M}\,[g_1]_n\right) \ ,
\\[0.5cm] &&
[e]_n = [\tilde e]_n + \frac{m}{M}\,[f_1]_{n-1}\ ,
\\[0.5cm] &&
[f_T]_n = \frac{n-1}{n}\,[\tilde f_T]_n\ ,
\\ &&
[f_{1T}^{\perp(1)}]_n = \frac{1}{n+1}[\tilde f_T]_{n+1}\ ,
\\[0.5cm] &&
[h]_n = \frac{n-1}{n+1}\,[\tilde h]_n\ ,
\\ &&
[h_{1}^{\perp (1)}]_n = \frac{1}{n+2}\,[\tilde h]_{n+1} \ ,
\\[0.5cm] &&
[e_L]_n = [\tilde e_L]_n\ .
\eea
Actually, we need not consider the five T-odd functions separately. They can be
simply considered as imaginary parts of other functions, when we allow complex
functions. In particular one can expand the correlation functions into
matrices in Dirac space~\cite{BBHM} to show that the relevant combinations are 
$(g_{1T} - i\,f_{1T}^{\perp})$ which we can treat together as one complex 
function $g_{1T}$. Similarly we can use complex functions
$(h_{1L}^\perp + i\,h_1^\perp)$ $\rightarrow$ $h_{1L}^\perp$,
$(g_T + i\,f_T)$ $\rightarrow$ $g_T$,
$(h_L + i\,h)$ $\rightarrow$ $h_L$,
$(e + i\,e_L)$ $\rightarrow$ $e$.

\section{Evolution equations}
In case of autonomous evolution of a function $f$ one has 
\be
\frac{d}{d\tau} \,f(x,\tau) = \frac{\alpha_s(\tau)}{2\pi}
\,\int_x^1 \frac{dy}{y} \ P^{[f]}\left(\frac{x}{y}\right)\,f(y,\tau),
\ee
where $\tau$ = $\ln Q^2$ and $P^{[f]}$ are the splitting functions. Using  
moments $A_n^{[f]}$ of these splitting functions \-- the anomalous
dimensions \--, this is 
\be
\frac{d}{d\tau} \,[f]_n(\tau) =
\frac{\alpha_s(\tau)}{2\pi}\,A_n^{[f]}\,[f]_n(\tau). 
\ee
This applies to 
the leading order results for the non-singlet twist-2 functions (with the
usual + prescription) \cite{AP,Baldra81},
\bea
&&
P^{[f_1]}(\beta) = P^{[g_1]}(\beta) = C_F\left[\frac{3}{2}\,\delta(1-\beta) +
\frac{1+\beta^2}{(1-\beta)_+}\right], 
\\ &&
P^{[h_1]}(\beta) = C_F\left[\frac{3}{2}\,\delta(1-\beta) +
\frac{2\beta}{(1-\beta)_+}\right], 
\eea
and the large $N_c$ result in leading order 
for the interaction-dependent functions \cite{ABH}
\bea
&&
P^{[\tilde f]}(\beta) =  \frac{N_c}{2}\left[
\frac{1}{2}\,\delta(1-\beta) + \frac{2}{(1-\beta)_+} + c\right],
\eea
with $c = -1$ for $\tilde g_T$, $c = -3$ for $\tilde h_L$ and $c=+1$ for
$\tilde e$. 
The corresponding anomalous dimensions are
\bea
&& A_n^{[f_1]} = A_n^{[g_1]} = C_F\left[
\frac{3}{2} + \frac{1}{n(n+1)} - 2\sum_{j=1}^n \frac{1}{j}\right] ,
\\
&& A_n^{[h_1]} = C_F\left[
\frac{3}{2} - 2\sum_{j=1}^n \frac{1}{j} \right],
\eea
and for the interaction-dependent functions in the large $N_c$ limit
\bea
&& A_n^{[\tilde g_T]} = \frac{N_c}{2}\left[
\frac{1}{2} + \frac{1}{n} - 2\sum_{j=1}^n \frac{1}{j} \right],
\label{DF31}
\\
&& A_n^{[\tilde h_L]} = \frac{N_c}{2}\left[
\frac{1}{2} - \frac{1}{n} - 2\sum_{j=1}^n \frac{1}{j} \right],
\\
&& A_n^{[\tilde e]} = \frac{N_c}{2}\left[
\frac{1}{2} + \frac{3}{n} - 2\sum_{j=1}^n \frac{1}{j} \right].
\label{DF33}
\eea
Using the moment analysis of the previous section, it is straightforward to
find that the evolution of $g_{1T}^{(1)}$ is driven not only by this function
itself but also by a higher moment of $g_1$ and a similar situation for
$h_{1L}^{\perp (1)}$. In the large $N_c$ limit ($C_F \rightarrow N_c/2$)
one obtains (omitting mass terms) 
\bea
&&
\frac{d}{d\tau}\,[g_{1T}^{(1)}]_n = \frac{\alpha_s(\tau)}{4\pi}\,N_c\left\{
\left[\frac{1}{2} - \frac{1}{n+1} - 2\sum_{j=1}^n \frac{1}{j}\right]
[g_{1T}^{(1)}]_n + \frac{1}{n+2}\,[g_1]_{n+1} 
\right\} ,
\\ &&
\frac{d}{d\tau}\,[h_{1L}^{\perp(1)}]_n =\frac{\alpha_s(\tau)}{4\pi}\,N_c\left\{
\left[\frac{1}{2} - \frac{3}{n+1} - 2\sum_{j=1}^n \frac{1}{j}\right]
[h_{1L}^{\perp (1)}]_n - \frac{1}{n+1}\,[h_1]_{n+1} 
\right\} ,
\eea
or in terms of the functions of light-cone momentum fractions 
\bea
&&
\frac{d}{d\tau}\,g_{1T}^{(1)}(x,\tau) 
= \frac{\alpha_s(\tau)}{4\pi}\,N_c\int_x^1 dy\,\left\{
\left[\frac{1}{2}\,\delta(y-x) + \frac{x^2+xy}{y^2(y-x)_+}\right]
\,g_{1T}^{(1)}(y,\tau)
+ \frac{x^2}{y^2}\,g_1(y,\tau)\right\} ,
\\ && 
\frac{d}{d\tau}\,h_{1L}^{\perp (1)}(x,\tau) 
= \frac{\alpha_s(\tau)}{4\pi}\,N_c\int_x^1 dy\,\left\{
\left[\frac{1}{2}\,\delta(y-x) + \frac{3x^2-xy}{y^2(y-x)_+}\right]
\,h_{1L}^{\perp (1)}(y,\tau)
-\frac{x}{y}\,h_1(y,\tau)\right\}.
\eea
Next we note that apart from a
$\gamma_5$ matrix the operator structures of the T-odd functions 
$f_{1T}^{\perp (1)}$ and $h_1^{\perp (1)}$ are in fact the same 
as those of $g_{1T}^{(1)}$ and $h_{1L}^{\perp (1)}$ (they can be considered as
the imaginary part of these functions~\cite{BBHM}). This implies that for the
non-singlet functions, one immediately can obtain the evolution of the T-odd 
functions, 
\bea
&&
\frac{d}{d\tau}\,[f_{1T}^{\perp(1)}]_n =\frac{\alpha_s(\tau)}{4\pi}\,N_c
\left[\frac{1}{2} - \frac{1}{n+1} - 2\sum_{j=1}^n \frac{1}{j}\right]
[f_{1T}^{\perp(1)}]_n ,
\\ &&
\frac{d}{d\tau}\,[h_{1}^{\perp(1)}]_n =\frac{\alpha_s(\tau)}{4\pi}\,N_c
\left[\frac{1}{2} - \frac{3}{n+1} - 2\sum_{j=1}^n \frac{1}{j}\right]
[h_{1}^{\perp(1)}]_n .
\eea
Furthermore, for the chiral-odd functions, which do not mix with a gluon
distribution, there is no difference between the non-singlet and the singlet
evolution. 
 
In the large $N_c$ limit, the evolution equations for the non-singlet
T-odd functions are of simple diagonal form with splitting functions
\bea
&&
P^{[f_{1T}^{\perp (1)}]} (\beta) = \frac{N_c}{2}\left[
\frac{1}{2}\,\delta(1-\beta) + \frac{\beta+\beta^2}{(1-\beta)_+} \right],
\\ &&
P^{[h_{1}^{\perp (1)}]} (\beta) =\frac{N_c}{2}\left[
\frac{1}{2}\,\delta(1-\beta) + \frac{3\beta^2-\beta}{(1-\beta)_+} \right].
\eea
Actually, we also obtain the anomalous dimensions (and splitting
functions) of the
T-odd twist-3 functions using $A^{[f_T]}$ = $A^{[\tilde f_T]}$ = $A^{[\tilde
g_T]}$, $A^{[h]}$ = $A^{[\tilde h]}$ = $A^{[\tilde h_L]}$ and
$A^{[e_L]}$ = $A^{[\tilde e_L]}$ = $A^{[e]}$ = $A^{[\tilde e]}$.

\section{Fragmentation functions}

Just as for the distribution functions one can perform an analysis of
the soft part describing quark fragmentation.
One needs~\cite{CS82} 
\be
\Delta_{ij}(z,\bm k_\st) =
\left. \sum_X \int \frac{d\xi^+d^2\bm \xi_\st}{(2\pi)^3} \,
e^{ik\cdot \xi} \,Tr  \langle 0 \vert {\cal U}(\infty,\xi)\,\psi_i (\xi) \vert
P_h,X\rangle \langle P_h,X\vert\overline \psi_j(0)\,{\cal U}(0,\infty) \vert 0
\rangle \right|_{\xi^- = 0}.
\ee
Note that because of the definition of the 
light-like vectors $n_\pm$ (via $P$ and $P_h$), the role of these vectors
for fragmentation functions will be interchanged with respect to the
distribution functions.
For the production of unpolarized or spin-1/2 hadrons $h$ in semi-inclusive 
hard scattering processes
one needs to leading order in $1/Q$ the correlation function \cite{MT96}
\begin{eqnarray}
\Delta(z,\bm{k}_\st) & = & 
zD_1(z ,\bm k^{\prime 2}_\st)\,\slsh n_- + 
zD_{1T}^\perp(z ,\bm k^{\prime 2}_\st)\, \frac{\epsilon_{\mu \nu \rho
\sigma}\gamma^\mu  n_-^\nu k_\st^\rho S_{h\st}^\sigma}{M_h}
\nonumber \\ && \mbox{}  
- zG_{1s}(z ,\bm k^{\prime}_\st)\, \slsh n_- \gamma_5
+ zH_{1T}(z ,\bm k^{\prime 2}_\st)\,\frac{[\slsh S_{h\st} ,
\slsh n_- ]\,\gamma_5}{2}
\nonumber \\ && \mbox{}
+ zH_{1s}^\perp(z,\bm k^{\prime}_\st)\,\frac{[\slsh k_\st ,
\slsh n_- ]\,\gamma_5}{2M_h} 
+ i\,zH_1^\perp (z,\bm k^{\prime 2}_\st) 
\,\frac{[ \slsh k_\st , \slsh n_- ]}{2M_h} 
+ {\cal O}\left(\frac{M_h}{P_h^-}\right).
\label{paramDeltazkt}
\end{eqnarray}
We used the shorthand notation
\beq
G_{1s}(z, \bm k_\st) \equiv
S_{h\sL}\,G_{1L}(z ,\bm k^{\prime 2}_\st)
+ \frac{(\bkt\cdot\bm{S}_{h\st})}{M_h}\,G_{1T}(z ,\bm k^{\prime 2}_\st) ,
\eeq
etc.
The arguments of the fragmentation functions are
$z$ = $P_h^-/k^-$ and $\bm k^\prime_\st$ = $-z\bm k_\st$. The first
is the (lightcone) momentum fraction of the produced hadron, the second
is the transverse momentum of the produced hadron with respect to the quark.
The $\bm k_\st$-integrated results are, using $F(z) \equiv \int d^2
k^\prime_\st\ F(z,\bm k_\st^{\prime 2})$ and 
$F^{(n)}(z) \equiv \int d^2k_\st^\prime\,(\bm k_\st^2/2M_h^2)^n
\,F(z,\bm k^{\prime 2}_\st)$ 
= $\int d^2k_\st^\prime\,(\bm k^{\prime 2}_\st/2z^2M_h^2)^n
\,F(z,\bm k^{\prime 2}_\st)$,
\bea
\Delta^{{\rm twist}-2}(z) &=& \frac{D_1(z)}{z} \slsh n_- 
+ S_{h\sL}\,\frac{G_1(z)}{z}\,\gamma_{5}\slsh n_-
+ \frac{H_1(z)}{z}\,\gamma_{5}\slsh S_{h\st}\slsh n_- ,
\\ 
\Delta^{{\rm twist}-3}(x) &=& \frac{M_h}{P_h^-} \left\{\frac{E(z)}{z} 
+ \frac{G_T(z)}{z}\,\gamma_{5}\slsh S_{h\st}
+ S_{h\sL}\,\frac{H_L(z)}{z}\,\gamma_{5}\frac{[\slsh n_-,\slsh n_+]}{2}
\right\}  
\nonumber \\ &&\quad \mbox{} 
+ \frac{M_h}{P_h^-} \left\{-i\,S_{h\sL}\,\frac{E_L(z)}{z}
\gamma_5 - \frac{D_T(z)}{z}\,\epsilon_\st^{\rho\sigma}\gamma_\rho
S_{h\st\sigma} + i\,\frac{H(z)}{z}\,\frac{[\slsh n_-,\slsh n_+]}{2} \right\}, 
\\
\Delta_\partial^\alpha (z) & = &
-\,\frac{G_{1T}^{(1)}(z)}{z}\,S_{h\st}^\alpha\,\slsh n_-\gamma_5
-S_{h\sL}\,\frac{H_{1L}^{\perp (1)}(z)}{z}
\,\frac{[\gamma^\alpha,\slsh n_-]\gamma_5}{2}
\nonumber \\ &&\quad \mbox{}
-\,\frac{{D_{1T}^{\perp (1)}}(z)}{z}
\,\epsilon^{\alpha}_{\ \ \mu\nu\rho}\gamma^\mu n_+^\nu {S_{h\st}^\rho}
- i\,\frac{{H_1^{\perp (1)}}(z)}{z}
\,\frac{[\gamma^\alpha, \slsh n_-]}{2}.
\label{Deltad}
\eea
In the twist-3 functions one can again isolate the interaction-dependent parts
as done for the distribution functions. They are now given by
\bea
&&
\tilde G_T(z) = G_T(z) - z\,G_{1T}^{(1)}(z) - \frac{m}{M_h}\,z H_1(z),
\\ &&
\tilde H_L(z) = H_L(z) + 2z\,H_{1L}^{\perp (1)}(z) - \frac{m}{M_h}\,z G_1(z),
\\ &&
\tilde E(z) = E(z) - \frac{m}{M_h}\,zD_1(z),
\\ &&
\tilde D_T(z) = D_T(z) + z\,D_{1T}^{\perp (1)}(z) ,
\\ &&
\tilde H(z) = H(z) + 2z\,H_1^{\perp (1)}(z),
\\ &&
\tilde E_L(z) = E_L(z).
\eea
For the $\bm k_\st$-integrated or the $\bm k_\st^2/2M_h$-weighted
fragmentation functions all results are obtained from the distribution 
functions by replacing 
$x \rightarrow 1/z$ and 
$f_{\ldots} (x) \longrightarrow D_{\ldots}(z)/z$, $g_{\ldots} (x) \rightarrow
G_{\ldots}(z)/z$ and $h_{\ldots} (x) \rightarrow H_{\ldots}(z)/z$.
The same applies to the relations from Lorentz invariance \cite{MT96,JM97}
\bea
&&G_T(z) = G_1(z) - z^3\frac{d}{dz}\left[
\frac{G_{1T}^{(1)}(z)}{z}\right],\\
&&H_L(z) = H_1(z) + z^3\frac{d}{dz}
\left[ \frac{H_{1L}^{\perp (1)}(z)}{z}\right],\\
&&D_T(z) = 
z^3\frac{d}{dz} \left[ \frac{D_{1T}^{\perp(1)}(z)}{z}\right],\\
&&H(z) = z^3\frac{d}{dz} \left[ \frac{H_1^{\perp (1)}(z)}{z}\right].
\eea
Expressing the functions in twist-2 functions and interaction-dependent
functions gives
\bea
&&
\frac{G_T(z)}{z} = 
-\int_z^1 dy\ \frac{G_1(y)}{y^2}
+ \frac{m}{M_h}\left[ H_1(z) + \int_z^1 dy\ \frac{H_1(y)}{y}\right]
+ \left[ \frac{\tilde G_T(z)}{z} + \int_z^1 dy\ \frac{\tilde
G_T(y)}{y^2}\right], 
\\ &&
G_{1T}^{(1)}(z) =
-\int_z^1 dy\ \frac{G_1(y)}{y^2}
+ \frac{m}{M_h}\int_z^1 dy\ \frac{H_1(y)}{y}
+ \int_z^1 dy\ \frac{\tilde G_T(y)}{y^2},
\\[0.5cm] &&
H_L(z) = 
-2\int_z^1 dy\ \frac{H_1(y)}{y}
+ \frac{m}{M_h}\left[ zG_1(z) + 2\int_z^1 dy\ G_1(y)\right]
+ \left[ \tilde H_L(z) + 2\int_z^1 dy\ \frac{\tilde H_L(y)}{y}\right],
\\ &&
z\,H_{1L}^{\perp(1)}(z) =
\int_z^1 dy\ \frac{H_1(y)}{y}
- \frac{m}{M_h}\int_z^1 dy\ G_1(y)
- \int_z^1 dy\ \frac{\tilde H_L(y)}{y},
\\[0.5cm] &&
E(z) =
\tilde E(z) + \frac{m}{M_h}\,zD_1(z),
\\[0.5cm] &&
\frac{D_T(z)}{z} = 
\left[ \frac{\tilde D_T(z)}{z} + \int_z^1 dy\ \frac{\tilde
D_T(y)}{y^2}\right] , 
\\ &&
D_{1T}^{\perp (1)}(z) =
- \int_z^1 dy\ \frac{\tilde D_T(y)}{y^2} ,
\\[0.5cm] &&
H(z) = 
\left[ \tilde H(z) + 2\int_z^1 dy\ \frac{\tilde H(y)}{y}\right],
\\ &&
z\,H_{1}^{\perp(1)}(z) =
-\int_z^1 dy\ \frac{\tilde H(y)}{y} ,
\\[0.5cm] &&
E_L(z) =
\tilde E_L(z).
\eea
The relations for the moments of fragmentation functions can be obtained 
from the above equations or from the results of the distribution 
functions via the replacements $n \rightarrow -n$ followed by $[f]_{-n}
\rightarrow [D/z]_n = [D]_{n-1}$. This yields  
\bea
&&
[G_T]_n = -\,\frac{1}{n+1}\,[G_1]_n 
+ \frac{n+2}{n+1}\,[\tilde G_T]_n
+ \frac{m}{M_h}\,\frac{n+2}{n+1}\,[H_1]_{n+1} \ ,
\\ &&
[G_{1T}^{(1)}]_{n+1} = -\,\frac{1}{n+1}\left(
[G_1]_{n} - [\tilde G_T]_{n} - \frac{m}{M_h}\,[H_1]_{n+1}\right) \ ,
\\[0.5cm] &&
[H_L]_n = -\,\frac{2}{n}\,[H_1]_n 
+ \frac{n+2}{n}\,[\tilde H_L]_n
+ \frac{m}{M_h}\,\frac{n+2}{n}\,[G_1]_{n+1} \ ,
\\ &&
[H_{1L}^{\perp (1)}]_{n+1} = \frac{1}{n}\left(
[H_1]_{n} - [\tilde H_L]_{n} - \frac{m}{M_h}\,[G_1]_{n+1}\right) \ ,
\\[0.5cm] &&
[E]_n = [\tilde E]_n + \frac{m}{M_h}\,[D_1]_{n+1}\ ,
\\[0.5cm] &&
[D_T]_n = \frac{n+2}{n+1}\,[\tilde D_T]_n\ ,
\\ &&
[D_{1T}^{\perp(1)}]_{n+1} = -\,\frac{1}{n+1}\,[\tilde D_T]_{n}\ ,
\\[0.5cm] &&
[H]_n = \frac{n+2}{n}\,[\tilde H]_n\ ,
\\ &&
[H_{1}^{\perp (1)}]_{n+1} = -\,\frac{1}{n}\,[\tilde H]_{n} \ ,
\\[0.5cm] &&
[E_L]_n = [\tilde E_L]_n\ .
\eea
The autonomous evolution equations are again of the form  
\be
\frac{d}{d\tau} \,D(z,\tau) = \frac{\alpha_s(\tau)}{2\pi}\int_z^1 \frac{dy}{y}
\ P^{[D]}\left(\frac{z}{y}\right)\,D(y,\tau),
\ee
or via the (usual) moments $A_n^{[D]} = \int_0^1 dz\ z^{n-1}\,P^{[D]}(z)$ 
of the splitting functions, 
\be
\frac{d}{d\tau} \,[D]_n(\tau) = \frac{\alpha_s(\tau)}{2\pi}
\,A_n^{[D]}\,[D]_n(\tau).
\ee
For the leading order contributions the analytic structure of the 
corrections for fragmentation functions is similar as for distribution
functions. We note a (generalized) Gribov-Lipatov reciprocity,
summarized by the following procedure. 
The splitting functions for
distribution functions $f(x,\tau)$ and corresponding fragmentation functions
$z\,D(z,\tau)$ are related by 
\bea
&&P^{[f]}(\beta) = \frac{{\cal N}(\beta)}{(1-\beta)_+},
\label{sff}
\\
&&P^{[zD]}(\beta) = \frac{\beta^2\,{\cal N}(1/\beta)}{(1-\beta)_+} .
\label{sfD}
\eea
This relation works for the twist-2 fragmentation functions {\em and} the
interaction-dependent functions~\cite{Belitsky-97a}, for ${\cal N}(\beta)$ 
being (at most a quadratic)
polynomial in $\beta$. In the case of the twist-2 functions the functional
form of the splitting functions is the same for distribution and
fragmentation functions. This is no longer true for the interaction-dependent
functions.  For the anomalous dimensions of
distribution and fragmentation functions the relation becomes 
\bea
&&
A_n^{[f]} = {\cal A}(n) -2\sum_{j=1}^{n} \frac{1}{j}
= {\cal A}(n) -2\gamma_E - 2\psi(n+1) ,
\label{reciprocity-1}
\\ &&
A_{n+1}^{[D]} = {\cal A}\left(-(n+1)\right)-2\gamma_E - 2\psi(n+1)
= {\cal A}\left(-(n+1)\right) -2\sum_{j=1}^n \frac{1}{j},
\label{reciprocity-2}
\eea
where ${\cal A}(n)$ is a rational function.  
We have not yet investigated the wider applicability of the above relations.
We find for the twist-2 fragmentation functions the familiar results, which 
obey the
original Gribov-Lipatov reciprocity relation $A_n^{[f]}=A_{n+1}^{[D]}$
between the twist-2 distribution functions $f = {f_1,g_1,h_1}$ and
fragmentation functions  $D = {D_1,G_1,H_1}$,
\bea
&&
A_{n+1}^{[D_1]} = A_{n+1}^{[G_1]} 
= C_F\left[\frac{3}{2} + \frac{1}{n(n+1)} -2\sum_{j=1}^n \frac{1}{j}\right] , 
\\ &&
A_{n+1}^{[H_1]} 
= C_F\left[\frac{3}{2} -2\sum_{j=1}^n \frac{1}{j}\right] . 
\eea
In the large $N_c$ limit, our generalized reciprocity relations in Eqs.\
(\ref{reciprocity-1}) and (\ref{reciprocity-2}) applied to Eqs.\ (\ref{DF31}) 
- (\ref{DF33}) give the results for the
interaction-dependent functions~\cite{Belitsky-97a,Belitsky-97b} 
\bea
&& A_{n+1}^{[\tilde G_T]} = \frac{N_c}{2}\left[
\frac{1}{2} - \frac{1}{n+1} - 2\sum_{j=1}^n \frac{1}{j} \right],
\\
&& A_{n+1}^{[\tilde H_L]} = \frac{N_c}{2}\left[
\frac{1}{2} + \frac{1}{n+1} - 2\sum_{j=1}^n \frac{1}{j} \right],
\\
&& A_{n+1}^{[\tilde E]} = \frac{N_c}{2}\left[
\frac{1}{2} - \frac{3}{n+1} - 2\sum_{j=1}^n \frac{1}{j} \right].
\eea
Again one then also knows $A^{[D_T]}$ = $A^{[\tilde D_T]}$ = $A^{[\tilde
G_T]}$, $A^{[H]}$ = $A^{[\tilde H]}$ = $A^{[\tilde H_L]}$ and
$A^{[E_L]}$ = $A^{[\tilde E_L]}$ = $A^{[E]}$ = $A^{[\tilde E]}$.

Using the moment analysis (the reciprocity relations cannot be used
straightforwardly) one obtains, omitting the mass terms, 
\bea
&&
\frac{d}{d\tau}\,[G_{1T}^{(1)}]_{n+1} = \frac{\alpha_s(\tau)}{4\pi}\,N_c\left\{
\left[\frac{1}{2} + \frac{1}{n} - 2\sum_{j=1}^n \frac{1}{j}\right]
[G_{1T}^{(1)}]_{n+1} - \frac{n}{(n-1)(n+1)}\,[G_1]_{n} 
\right\} ,
\\ &&
\frac{d}{d\tau}\,[H_{1L}^{\perp(1)}]_{n+1} =\frac{\alpha_s(\tau)}{4\pi}
\,N_c\left\{
\left[\frac{1}{2} + \frac{3}{n} - 2\sum_{j=1}^n \frac{1}{j}\right]
[H_{1L}^{\perp (1)}]_{n+1} + \frac{n-1}{n^2}\,[H_1]_{n} 
\right\} ,
\eea
with in this case mixing with a lower moment of the twist-2 functions.   
In terms of the functions of lightcone momentum fractions one finds
\bea
&&
\frac{d}{d\tau}\,zG_{1T}^{(1)}(z,\tau) 
= \frac{\alpha_s(\tau)}{4\pi}\,N_c\int_z^1 dy\,\left\{
\left[\frac{1}{2}\,\delta(y-z) + \frac{y+z}{y(y-z)_+}\right]
\,yG_{1T}^{(1)}(y,\tau)
-\frac{y^2 + z^2}{2y^2 z}\,G_1(y,\tau)\right\} ,
\\ && 
\frac{d}{d\tau}\,zH_{1L}^{\perp (1)}(z,\tau) 
= \frac{\alpha_s(\tau)}{4\pi}\,N_c\int_z^1 dy\,\left\{
\left[\frac{1}{2}\,\delta(y-z) + \frac{3y-z}{y(y-z)_+}\right]
\,yH_{1L}^{\perp (1)}(y,\tau)
+\frac{1 + \ln(z/y)}{y}\,H_1(y,\tau)\right\}.
\eea
Given the fact that, apart from an additional $\gamma_5$, the operator 
structure for the T-odd Sivers and
Collins functions, $D_{1T}^{\perp (1)}$ and  $H_1^{\perp (1)}$, are the same as
those of $G_{1T}^{(1)}$ and $H_{1L}^{\perp (1)}$ but without mixing with 
$G_1$ or $H_1$, one finds in the large $N_c$ limit an autonomous
evolution for the T-odd functions, with anomalous dimensions  
\bea
&&
A_{n+1}^{[D_{1T}^{\perp(1)}]} =\frac{N_c}{2}\left[
\frac{1}{2} + \frac{1}{n} - 2\sum_{j=1}^n \frac{1}{j}\right],
\label{A124}
\\ &&
A_{n+1}^{[H_{1}^{\perp(1)}]} =\frac{N_c}{2}\left[
\frac{1}{2} + \frac{3}{n} - 2\sum_{j=1}^n \frac{1}{j}\right].
\label{A125}
\eea
corresponding to splitting functions
\bea
&&
P^{[zD_{1T}^{\perp (1)}]} (\beta) = \frac{N_c}{2}\left[
\frac{1}{2}\,\delta(1-\beta) + \frac{1+\beta}{(1-\beta)_+} \right],
\label{P126}
\\ &&
P^{[zH_{1}^{\perp (1)}]} (\beta) =\frac{N_c}{2}\left[
\frac{1}{2}\,\delta(1-\beta) + \frac{3-\beta}{(1-\beta)_+} \right].
\label{P127}
\eea
The results Eqs.\ (\ref{A125}) and (\ref{P127}) should prove useful for 
studies of the Collins effect and Eqs.\ (\ref{A124}) and (\ref{P126}) for 
studies of transversely polarized $\Lambda$ production \cite{ABDM-01}. 

\begin{center}
\begin{table}
\caption{\label{moments}
The anomalous dimensions from which the large $Q^2$
behavior of the moments, proportional to $[\alpha_s(Q^2)]^{d_n}$, is obtained.
Defining the moments $a_n$ taking out the factor $C_F$ or $N_c/2$ from the
anomalous dimensions $A_n$, one has for the twist-2 functions
$d_n = - 2a_n\,C_F/b_0$ with $b_0 = (11\,N_c - 2\,N_f)/3$,
while for the large
$N_c$ results one has $d_n = - 3\,a_n/11$. Also indicated is the charge 
conjugation behavior of the functions, $\bar f(x) = \pm f(-x)$.}  
\begin{tabular}{l|c|ccc|l}
function & \mbox{}\qquad C\qquad\mbox{}& $a_1$ & $a_2$ &
\mbox{}\qquad $a_3$\qquad\mbox{} & validity\\ 
\hline
$f_1$ & $-$ & 0 & -7/6 & -25/12 & \\
$g_1$ &  + & 0 & -7/6 & -25/12 & \\
$h_1$ & $-$  & -1/2 & -3/2 & -13/6 & \\
\hline
$\tilde g_{T}$ and $\tilde f_T$ & + & -1/2 &-2 & -17/6 & large $N_c$\\  
$\tilde h_{L}$ and $\tilde h$ & $-$ & -5/2 & -3 & -7/2 & large $N_c$\\ 
$\tilde e$  & + & +3/2 & -1 & -13/6 & large $N_c$\\ 
\hline
$g_{1T}^{(1)}$ and $f_{1T}^{\perp (1)}$ 
& $-$  & -2 &-17/6 & -41/12 & large $N_c$\\  
$h_{1L}^{\perp(1)}$ and $h_1^{\perp(1)}$  
& +  & -3 & -7/2 & -47/12 & large $N_c$\\ 
\hline
$zD_1$ & $-$ & 0 & -7/6 & -25/12 \\ 
$zG_1$ &  +  & 0 & -7/6 & -25/12 \\ 
$zH_1$  & $-$  & -1/2 & -3/2 & -13/6 & \\
\hline
$z\tilde G_{T}$ and $z\tilde D_T$ & +  & -2 & -17/6 & -41/12 & large $N_c$\\  
$z\tilde H_{L}$ and $z\tilde H$ & $-$  & -1 & -13/6 & -35/12 & large $N_c$\\ 
$z\tilde E$  & +  & -3 & -7/2 & -47/12 & large $N_c$\\ 
\hline
$zG_{1T}^{(1)}$ and $zD_{1T}^{\perp (1)}$ 
&  $-$ & -1/2 &-2 & -17/6 & large $N_c$\\  
$zH_{1L}^{\perp(1)}$ and $zH_1^{\perp(1)}$ 
& + &  +3/2 & -1 & -13/6 & large $N_c$
\end{tabular}
\end{table}
\end{center}

\section{Discussion and conclusions}

Our goal was to obtain the evolution equations of the functions that appear in
azimuthal spin asymmetries. These $p_\st$-dependent
functions appear in asymmetries that are not suppressed by explicit powers of
the hard momentum. But as functions of transverse momentum they are not of 
definite twist, which implies that in order to obtain the evolution equations 
one has to calculate corrections to higher twist operators as well. 
For the first 
$\bm p_\st^2/2M^2$ moment (transverse moment) of these $p_\st$-dependent
functions, such as for the Collins fragmentation function $H_1^{\perp (1)} =
\int  d^2k_\st^\prime\,\bm k^{\prime 2}_\st/2z^2M_h^2\,H_1^\perp(z,\bm
k^{\prime 2}_\st)$, we obtain DGLAP-like evolution equations.
Such moments appear in cross sections weighted with the
momentum $q_\st^\alpha$,  where only the directional (azimuthal) dependence
remains. For explicit examples we refer to 
Refs.~\cite{BJM-97,Boer-Mulders-98}.  In case one does not weight 
the transverse momentum integration of the differential cross section, one is 
only sensitive to the twist-2
functions $f_1, g_1$ and $h_1$ (and their fragmentation counterparts), but in 
case one weighs with one or more powers of the observed transverse momentum, 
one becomes sensitive to the functions $g_{1T}^{(1)}, h_{1L}^{\perp\, (1)}, 
f_{1T}^{\perp\, (1)}, h_1^{\perp\, (1)}$ (and their fragmentation 
counterparts), which are functions of the lightcone momentum fraction $x$ only.

In the large-$N_c$ limit, the non-singlet evolution of 
these functions involves only the functions themselves and (in the
T-even case) only well-known twist-2 functions. For the chiral-odd functions 
the equations also apply to the singlet case, since there is no mixing with
gluon distribution functions. The large-$N_c$ evolution
equations are expected to be good approximations to the full evolution
equations which are not of this simple form (cf.\ Ref.\ \cite{Beyond-LO-Nc}),
because of the appearance of two-argument twist-3 functions as in Eq.\ 
(\ref{twoarg}). It is not
excluded that the first $1/N_c$ correction to the result obtained here may
still lead to autonomous evolution equations, but we will not address this 
issue here. 
Especially the (large $N_c$) evolution equation we have obtained for
$H_1^{\perp (1)}$,
\be
\frac{d}{d\tau} \,z H_{1}^{\perp (1)}(z,\tau) 
= \frac{\alpha_s}{4\pi}\; N_c\; \int_z^1 dy\, \left[
\frac{1}{2}\,\delta(y-z) + \frac{3y-z}{y(y-z)_+} \right] 
\,y H_{1}^{\perp (1)}(y,\tau),
\ee
should prove  useful for the comparison of data on Collins
function asymmetries from  different experiments, performed at different
energies.

It is worth investigating the large $Q$ behavior of the solutions to 
the various evolution equations. 
For this purpose we have given the first 3 anomalous dimensions
for the different functions in Table~\ref{moments}. First we note that 
all (diagonal) anomalous dimensions of $g_{1T}^{(1)}, h_{1L}^{\perp\, (1)}, 
f_{1T}^{\perp\, (1)}$ and $h_1^{\perp\, (1)}$ are negative, implying that
these functions will vanish asymptotically ($Q^2 \to \infty$), except that 
for the T-even functions there is mixing with $g_1$ and $h_1$, but this does 
not alter the conclusion.

For the fragmentation counterparts the conclusion is similar, except for the
fact that the lowest anomalous dimensions of $zH_{1L}^{\perp (1)}$ and 
$zH_{1}^{\perp (1)}$ are positive, potentially leading to divergent behavior of
the functions as $Q^2 \to \infty$. However, here we recall the 
Sch\"afer-Teryaev sum rule \cite{ST-00}
\beq
\int dz\ z H_{1}^{\perp (1)}(z) = 0,
\eeq
which states that the first moment of $zH_{1}^{\perp (1)}$ is zero, making the
sign of the first anomalous dimension irrelevant. Similar sum rules hold for
the other first transverse moments of fragmentation 
functions~\cite{ST-00,ABDM-01}. All higher moments will vanish asymptotically. 
The behavior of the sum rule for the first moment of the
function $e$ is discussed in  Ref.~\cite{Jaffe-Ji-92}.

In conclusion, using Lorentz invariance and the QCD equations of motion, the
operator structure of the transverse moments of $p_\st$-dependent quark
distribution and fragmentation functions can be found in terms of twist-2 and
twist-3 operators. Knowing their, for large $N_c$ simple, evolution
one also knows the evolution of azimuthal asymmetries in semi-inclusive 
hard scattering processes. 

\acknowledgments 
We acknowledge early work of Rik Tangerman on this subject and
we thank Andrei Belitsky, Bob Jaffe, Werner Vogelsang for helpful comments and
discussions. D.B.\ thanks RIKEN, Brookhaven National Laboratory and the U.S. 
Department of Energy for support under contract \# DE-AC02-98CH10886.

\end{document}